\documentclass[prd,nofootinbib,tightenlines,preprint]{revtex4}
\usepackage{epsfig}
\usepackage{graphicx}
\usepackage{bm}

\def\bn{{\bm{n}}}

\def\ihat{\hat{\bm{\imath}}}
\def\Tr{\textrm{Tr}\,}

\def\nnn{{\it nnn}}

\def\Eq#1{Eq.~(\ref{#1})}

\begin{document}
\title{Spontaneous symmetry breaking in strong-coupling lattice QCD at high density}
\author{Barak Bringoltz}
\author{Benjamin Svetitsky}
\affiliation{School of Physics and Astronomy, Raymond and Beverly
Sackler Faculty of Exact Sciences, Tel Aviv University, 69978 Tel
Aviv, Israel}
\begin{abstract}
We determine the patterns of spontaneous symmetry breaking in
strong-coupling lattice QCD in a fixed background baryon density.
We employ a next-nearest-neighbor fermion formulation that
possesses the $SU(N_f)\times SU(N_f)$ chiral symmetry of the
continuum theory.  We find that the global symmetry of the ground state varies with
$N_f$ and with the background baryon density.  In all cases the
condensate breaks the discrete rotational symmetry of the lattice
as well as part of the chiral symmetry group.

\end{abstract}
 \maketitle

\section{Introduction}

The study of quantum chromodynamics at high baryon density, nearly
as old \cite{Collins} as the theory itself, has gained impetus in
recent years with new interest in the idea of color
superconductivity (CSC) \cite{Barrois,Bailin,Alford,Rapp}.  Recent
work has led to a rich phase structure, including the
possibilities of color-flavor locking \cite{Alford1} and
crystalline superconductivity \cite{Alford2}.  For a review see
\cite{Krishna}.

All this work depends on effective theories derived from (or at
least motivated by) weak-coupling QCD.  The running coupling,
however, will become weak only at high densities; in fact it turns
out that reliable calculations demand extremely high densities
\cite{Son,Shuster}. If any of the predictions for moderate
densities are to be believed, they must be confirmed by
non-perturbative methods, or at least by models that incorporate
QCD's strong-coupling features.

In a previous paper \cite{paper1} we constructed a framework for
introducing baryons to lattice QCD at strong coupling. Though the
lattice theory is a way of defining QCD at the most fundamental
level, this process requires weak coupling; the strong-coupling
theory may be regarded as an effective theory at large distances.
It displays the non-perturbative effects of color confinement and
spontaneous breakdown of chiral symmetry (for suitably chosen
fermion formulations).  Our framework is based on the Hamiltonian
approach. We use strong coupling perturbation theory to write an
effective Hamiltonian for color singlet objects \cite{Smit,SDQW}.
At lowest order we get an antiferromagnetic Hamiltonian that
describes meson physics with a fixed baryon background
distribution. (Baryons move only at higher order.) We study this
Hamiltonian through its path integral, which takes the form of
a nonlinear $\sigma$ model.

The strong-coupling theory has no color degrees of freedom.  Its
properties can be determined through study of its global
symmetries. The global symmetry group of the action depends on the
formulation of the lattice fermions. For naive, nearest-neighbor
(NN) fermions the symmetry is $U(N)$ with
\begin{equation}
N=4N_f,
\end{equation}
realized on the quark spinors by combining the Dirac index with
the flavor index.  This too-large symmetry is indicative of the
doubling problem of naive fermions \cite{NN}. Adding longer-range
terms to the fermion kernel can reduce this artificial symmetry.
We add a next-nearest-nearest neighbor (NNN) term inspired by the
SLAC fermion formulation \cite{DWY,SDQW}.  In weak coupling, of
course, this theory is still doubled; in strong coupling, there
are no apparent ill effects of this doubling, and the $U(N)$
symmetry is broken to the $U(N_f)\times U(N_f)$ symmetry of
continuum QCD.\footnote{The axial $U(1)$ symmetry of the lattice
theory is of course not present in continuum QCD, but it is
inevitable on the lattice if we start with a local, chirally
symmetric fermion theory \cite{NN}. We can mend our effective
theory by hand, by adding new terms derived from an 't~Hooft
instanton vertex.}

\section{Effective Hamiltonian and $\sigma$ model}

In  \cite{paper1} we studied the NN theory only, and showed that
its global $U(N)$ symmetry is spontaneously broken to a subgroup
that depends on the baryon number.  In this paper we add the NNN
terms to the Hamiltonian and thus study a theory with the same
symmetry as the continuum theory.  The effective Hamiltonian in
strong coupling is \cite{paper1}
\begin{equation}
\label{Hqm} H=J_1 \sum_{\bn i ,\eta} Q_{\bn}^{\eta}
Q_{\bn+\ihat}^{\eta} + J_2 \sum_{\bn i ,\eta} Q_{\bn}^{\eta}
Q_{\bn+2\ihat}^{\eta} s_i^{\eta}.
\end{equation}
Here $Q_{\bn}^{\eta}$ are $U(N)$ charges at site $\bn$, with
$\eta=1,\ldots,N^2$. This Hamiltonian moves mesonic excitations
around the lattice, leaving the baryon density fixed. The states
at $\bn$ comprise a $U(N)$ representation whose Young tableau has
$N_c$ columns. The number of rows $m$ depends on the baryon number
$B$ at $\bn$ according to
\begin{equation}
m=B+2N_f.
\end{equation}
The sign factors $s^{\eta}_i$ are given by
\begin{equation}
\label{sign_factor} s^{\eta}_i=2\, \Tr M^{\eta} \alpha_i M^{\eta}
\alpha_i.
\end{equation}
Here $M^{\eta}$ are  the matrices of the fundamental
representation of the $U(N)$ algebra, normalized in the usual way,
$\Tr M^{\eta} M^{\eta'} = \frac12 \delta_{\eta \eta'}$. The
matrices $\alpha_i$ are the $4\times4$ Dirac matrices times the
unit matrix in flavor space.

The NN term in the Hamiltonian is that of a $U(N)$
antiferromagnet, while the NNN terms break the symmetry to
$U(N_f)_L\times U(N_f)_R$.
If one derives the fermion Hamiltonian by
truncating the SLAC Hamiltonian, then both couplings $J_1$ and
$J_2$ are positive, and $\hbox{$J_2=J_1/8$}$. If we argue,
however, that the strong-coupling Hamiltonian is derived by
block-spin transformations applied to a short-distance
Hamiltonian, then we cannot say much about the couplings that
appear in it.  We will assume that couplings in the effective
Hamiltonian fall off strongly with distance, that is, $0<J_2 \ll
J_1$.

We use spin-coherent states \cite{Sachdev} to write the partition
function for the effective Hamiltonian~(\ref{Hqm}).  This is the
path integral for a Euclidean nonlinear $\sigma$ model. The
$\sigma$ field at site $\bn$ is an $N\times N$ hermitian, unitary
matrix that represents an element of the coset space
$U(N)/[U(m)\times U(N-m)]$. It can be written as a unitary
rotation of the reference matrix~$\Lambda$,
\begin{equation}
\sigma_\bn=U_\bn \Lambda U^{\dag}_\bn \label{sigma_n},
\end{equation}
where
\begin{equation}
\Lambda = \left( \begin{array}{cc}
        \bm1_{m}     & 0 \\
        0 &     -\bm1_{N-m}        \end{array}     \right). \label{Lambda}
\end{equation}
and $U_\bn\in U(N)$.

The action of the $\sigma$ model is
\begin{equation}
\label{S} S=\frac{N_c}2\int d\tau\left[-\sum_{\bn} \Tr\Lambda
U^{\dag}_{\bn} \partial_\tau U_{\bn} +\frac {J_1}2\sum_{\bn i} \Tr
\left( \sigma_\bn \sigma_{\bn+\ihat} \right) + \frac{J_2}2
\sum_{\bn i} \Tr \left( \sigma_\bn \alpha_i \sigma_{\bn+2\ihat}
\alpha_i\right)  \right].
\end{equation}
The NN term is invariant under the global $U(N)$ transformation
$U\to VU$, or $\sigma\to V\sigma V^\dag$.  The NNN term is only
invariant if $V^\dag \alpha_i V=\alpha_i$ for all $i$.  This
restricts $V$ to the form
\begin{equation}
V=\exp
\left[i\left(\theta_V^a+\gamma_5\theta_A^a\right)\lambda^a\right],
\end{equation}
where $\lambda^a$ are flavor generators.  This is
a chiral transformation in $U(N_f)\times U(N_f)$.

The NNN term couples (discrete) rotational symmetry to the internal symmetry, {\it viz.}
\begin{equation}
\sigma_\bn \to R^{\dag} \sigma_{\bn'} R, \qquad \bn'={\cal R} \bn.
\label{rotattions}
\end{equation}
Here ${\cal R}$ is a $90^o$ lattice rotation and $R$ represents it according to
\begin{equation}
R=\exp \left[ i \frac{\pi}4 \left( \begin{array}{cc}
         \sigma_j & 0 \\
        0 & \sigma_j \end{array} \right) \right] \otimes {\bf 1}_{N_f}.
\label{R}
\end{equation}

\section{Nearest-neighbor theory}

The overall $N_c$ factor in \Eq{S} allows a systematic treatment
in orders of $1/N_c$.  In leading order, the ground state is found
by minimizing the action, which gives field configurations that
are $\tau$ independent and that minimize the interaction terms. For the NN
theory, minimizing the single link interaction
\begin{equation}
\label{site2} E=\frac{J_1}2 \Tr \sigma_1 \sigma_2
\end{equation}
allows us to construct the vacuum by placing $\sigma_1$ and
$\sigma_2$ on the even and odd sites.

We impose a uniform baryon density, $B_\bn=B>0$, on the effective
Hamiltonian by setting a fixed value of $m>2N_f$ on every site. To
minimize the single-link energy~(\ref{site2}) we first choose a
basis where
\begin{equation}
\sigma_1=\Lambda=\left( \begin{array}{cc}
\bm1_m&0\\
0&-\bm1_{N-m}\end{array} \right). \label{twospin1}
\end{equation}
The analysis in \cite{paper1} then shows that $\sigma_2$ can take
any value of the form
\begin{equation}
\sigma_2=\left( \begin{array}{cc}
\sigma^{(m)}     &       0       \\
0       &       \bm1_{N-m} \end{array} \right),
\end{equation}
where the $m\times m$ submatrix $\sigma^{(m)}$ can be chosen
freely in the submanifold $\hbox{$U(m)/[U(2m-N)\times U(N-m)]$}$
according to
\begin{equation}
\sigma^{(m)}=U^{(m)}\Lambda^{(m)}U^{(m)\dag},
\end{equation}
with
\begin{equation}
\Lambda^{(m)}=\left( \begin{array}{cc}
\bm1_{2m-N}&0\\
0&-\bm1_{N-m}
\end{array}\right)
\label{eq:Lm}
\end{equation}
and $U^{(m)}\in U(m)$. As mentioned, we construct a ground state
of the infinite lattice by replicating $\sigma_1$ and $\sigma_2$
on the even and odd sites of the lattice. Thus while all the
$\sigma_\bn$ on the even sites point to $\Lambda$, on the odd
sites each $\sigma_\bn$ wanders independently in the submanifold.
This classical ground state has a huge degeneracy, exponential in
the volume.

In Ref.~\onlinecite{paper1} we showed that the $O(1/N_c)$
fluctuations generate a ferromagnetic interaction among the odd sites,
causing them to align to a common value (``order from disorder''
\cite{odo}). This is a N\'eel structure, with two sublattices. The
even sites break $U(N)$ to $\hbox{$U(m)\times U(N-m)$}$ and then
the odd sites break the symmetry further to $\hbox{$U(2m-N)\times
U(N-m) \times U(N-m)$}$.

\section{next-nearest-neighbor theory}
\label{sec:nnn} Now we add the NNN interactions to the effective
action.  At the classical level, they do not by themselves remove
the classical degeneracy of the NN theory.  We have to
introduce the $O(1/N_c)$ fluctuations \textit{first} in order to
stabilize the N\'eel ground state of the NN Hamiltonian. Hence we
assume $1/{N_c}>J_2/J_1$, and treat the NNN interactions as a
perturbation that lifts part of the (global) degeneracy of the
$O(1/N_c)$ ground state.

We begin, then, by assuming a N\'eel ansatz that minimizes the NN
term in the action (\ref{S}). The NNN term acts within each of the
two sublattices.  Writing $\sigma_{e,o}$ for the sublattice
fields, the NNN contribution to the energy per $2\times 2 \times 2$ lattice cell is
\begin{equation}
E_{\nnn} = \frac{J_2}2 \sum_{a=e,o} \sum_i \Tr \left[ \sigma_a
\alpha_i \sigma_a \alpha_i    \right]. \label{eq:E}
\end{equation}
Here $\sigma_a$ is a global unitary rotation of the solution to the NN theory given by Eqs.~(\ref{twospin1})--(\ref{eq:Lm}).

We can find a lower bound for $E_{\nnn}$. Writing in each term $\Sigma_1^a=\sigma_a$
and $\Sigma_2^{ai}=\alpha_{a i} \sigma_a \alpha_{ai}^\dag$, we have
\begin{equation}
E_{\nnn}=\frac{J_2}2 \sum_{a i} \Tr
\Sigma_1^a\Sigma_2^{ai}.\label{minim}
\end{equation}
Note that $\Sigma_{1,2}$ are unitary rotations of the reference matrix $\Lambda$.
Each term in \Eq{minim} may be bounded from below by allowing these unitary
rotations to vary independently over the
entire $U(N)$ group. This is just
the minimization problem posed in \Eq{site2} above. The solution is
given by Eqs.~(\ref{twospin1})--(\ref{eq:Lm}), whence the bound
\begin{equation}
E_{\nnn}>3J_2(4m-3N). \label{bound}
\end{equation}

We minimize $E_{\nnn}$ by writing an ansatz for $\sigma_a$ that
saturates the lower bound. At this point we choose to work in a
basis where $\gamma_5$ is diagonal,
\begin{equation}
\gamma_5=\left( \begin{array}{cc}
    \bm1_{N/2} & 0   \\
    0 & -\bm1_{N/2}  \end{array} \right).
\end{equation}
Our ansatz is
\begin{eqnarray}
\sigma_e&=&U\Lambda_e U^\dag= U\left( \begin{array}{cc}
    \bm1_m & 0   \\
    0 & -\bm1_{N-m}  \end{array} \right)U^\dag,\nonumber\\[2pt]
\sigma_o&=&U\Lambda_o U^\dag= U\left( \begin{array}{ccc}
    \bm1_{2m-N} & 0 &0  \\
    0 & -\bm1_{N-m}&0\\
    0&0&\bm1_{N-m}  \end{array} \right)U^\dag. \label{ansatz}
\end{eqnarray}
(Note that $\Lambda_o$ is a rotation of $\Lambda_e$.) This is a global rotation (via $U$) of a
particular configuration that minimizes the NN energy, as seen
above.  We further suppose that $U$ takes the form
\begin{equation}
 U=\frac1{\sqrt{2}} \left( \begin{array}{cc}
    u & u   \\
    -u & u  \end{array} \right),
\label{eq:Uo}
\end{equation}
with $u\in U(N/2)$.

The minimization of the energy via the ansatz is a problem of
vacuum alignment \cite{Peskin}. We begin with the unperturbed NN
problem, where the symmetry group $G\equiv U(N)$ is broken to
$H\equiv \hbox{$U(2m-N)\times U(N-m)\times U(N-m)$}$. Our
reference vacuum is given by $\sigma_{e,o}=\Lambda_{e,o}$, giving
a specific alignment of $H$ as the invariance group of this
vacuum. We determine the rotation matrix $U$ [within the ansatz
(\ref{eq:Uo})] that minimizes the energy of the perturbation
$E_{\nnn}$, which we write as
\begin{equation}
\label{E_bar} \label{Ennn} E_{\nnn}=\frac{J_2}2 \sum_{a i} \Tr
\Lambda_a \bar{\alpha}_{i} \Lambda_a \bar{\alpha}_{i}^\dag,
\end{equation}
where $\bar{\alpha}_{i}=U^{\dag} \alpha_i U=\bar{\alpha}_{i}^\dag$.

We denote the generators of $H$ collectively as $T$ and the
remaining generators of $G$ as $X$.  The $T$ matrices commute with
both $\Lambda_e$ and $\Lambda_o$, while the $X$ matrices do not.
The symmetry-breaking term in the energy is $E_{\nnn}$, given by
\Eq{Ennn} in terms of the rotated Hermitian matrices
$\bar{\alpha}_{i}$. We project each
$\bar{\alpha}_{i}$ onto the $T$ and $X$ subspaces, giving the
decomposition
\begin{equation}
\bar{\alpha}_i = \bar{\alpha}_i^{T} + \bar{\alpha}_i^X.
\end{equation}
Using the invariance of $\Lambda_{e,o}$ under $T$,
\begin{equation}
\left[ \Lambda_a , \bar{\alpha}_i^T \right]=0,
\end{equation}
and the orthogonality of the $T$ and $X$ subspaces,
\begin{equation}
\Tr \left[ \bar{\alpha}_i^T \bar{\alpha}_i^X \right]=0,
\label{eq:relations}
\end{equation}
we have
\begin{equation}
E_{\nnn}=\frac{J_2}2 \sum_{i,a} \Tr \left[ \left( \bar{\alpha}_i^T
\right)^2 + \Lambda_a \bar{\alpha}_i^X \Lambda_a \bar{\alpha}_i^X
\right] \label{eq:Etx}
\end{equation}
Following the block form of $\Lambda_{e,o}$, we divide the broken
generators $X$ into 3 sets, denoting them as $X_a$ with $a=1,2,3$.
Their structures are respectively
\begin{equation}
\left( \begin{array}{ccc}
    0 & \tilde{X}_1 & 0 \\
    \tilde{X}_1^{\dag} & 0 & 0  \\
    0 & 0 & 0   \\  \end{array} \right) ,\quad
\left( \begin{array}{ccc}
    0 & 0 & \tilde{X}_2 \\
    0 & 0 & 0   \\
    \tilde{X}_2^{\dag} & 0 & 0  \\  \end{array} \right)
    ,\quad\textrm{and}\
\left( \begin{array}{ccc}
    0 & 0 & 0   \\
    0 & 0 & \tilde{X}_3 \\
    0 & \tilde{X}_3^{\dag} & 0  \\  \end{array} \right).
\end{equation}
Writing $\bar{\alpha}_i^X=\sum_a \bar{\alpha}_i^{X_a}$, the
following relations can be proved easily:
\begin{equation}\begin{array}{c @{\qquad} c}
\left[ \Lambda_1 , \bar{\alpha}_i^{X_1} \right]=0 & \left\{\Lambda_2 , \bar{\alpha}_i^{X_1} \right\}=0
\\[4pt]
\left\{ \Lambda_1 , \bar{\alpha}_i^{X_2} \right\}=0 & \left[\Lambda_2 , \bar{\alpha}_i^{X_2} \right]=0
\\[4pt]
\left\{ \Lambda_1 , \bar{\alpha}_i^{X_3} \right\}=0 &
\left\{\Lambda_2 , \bar{\alpha}_i^{X_3} \right\}=0.
\end{array}\label{eq:com}
\end{equation}
Using these together with the further orthogonality conditions,
\begin{equation}
\Tr \left[ \bar{\alpha}_i^{X_a} \bar{\alpha}_i^{X_b} \right]
=0,\qquad a \neq b, \label{eq:indep}
\end{equation}
we bring \Eq{eq:Etx} to the form
\begin{equation}
E_{\nnn}=J_2 \sum_i \Tr \left[ \left( \bar{\alpha}_i^{T} \right)^2
-\left( \bar{\alpha}_i^{X_3} \right)^2 \right]. \label{eq:E_TX3}
\end{equation}

The rotation $U$
given in \Eq{eq:Uo} saturates the lower
bound for the energy. We will proceed to prove this for the case
$m \ge 3N/4$. In view of  \Eq{eq:Uo} we can write $\bar{\alpha}_i$
in the form
\begin{equation}
\label{al_bar} \bar{\alpha}_i = \left( \begin{array}{cc}
                        0 & \bar{\sigma}_i \\
                        \bar{\sigma}_i & 0 \end{array} \right),
\end{equation}
where $\bar{\sigma}_i\equiv u^{\dag} \sigma_i u$. It is
straightforward to check that for $m \ge 3N/4$ we have
\begin{eqnarray}
\bar{\alpha}_i^{X_3} &=& 0, \label{al_X3_above} \\
\bar{\alpha}^T_i &=& \left( \begin{array}{cc}
                0 & \bar{\sigma}'_i \\
                \bar{\sigma}'^{\dag}_i &  0  \end{array} \right), \label{al_T_above}
\end{eqnarray}
with $\bar{\sigma}'_i$ composed of the first $(2m-3N/2)$ columns
of $\bar{\sigma}_i$,
\begin{equation}
\left( \bar{\sigma}'_i \right)_{pq} = \left\{ \begin{array}{ll}
\left( \bar{\sigma}_i \right)_{pq}
& \quad \textrm{for} \quad q=1,\dots,2m-3N/2, \\[2pt]
0 & \quad \textrm{else}. \end{array} \right.
\end{equation}
The energy is
\begin{equation}
\label{E_m_above} E_{\nnn}=2J_2 \sum_{i,pq} |(\bar{\sigma}'_i)_{pq}|^2
= 2J_2 \sum_{i} \sum_{q=1}^{2m-3N/2} \left( \bar{\sigma}^{\dag}_i
\bar{\sigma}_i \right)_{qq} = 3J_2 (4m-3N),
\end{equation}
which is exactly the lower bound.

According to
\Eq{E_m_above}, the bound is saturated for any $u \in U(N/2)$. 
Different choices of $u$ are {\em not\/} in general related by transformations
of the $U(N_f)\times U(N_f)$ symmetry group.
There is thus an accidental degeneracy of the vacuum when the NNN term is treated classically.
This degeneracy is presumably lifted by fluctuations.

The simplicity of this calculation depends on the assumption
$m\ge3N/4$. For $m < 3N/4$, both $\bar{\alpha}^{X_3}_i$ and
$\bar{\alpha}^T_i$ are nonzero. Moreover the index structure of
the projections is more complex.\footnote{An exception is the
$m=N/2$ case ($B=0$), which was solved in \cite{paper1}.} Therefore
in these cases we resort to numerical minimization of \Eq{Ennn}
over $u\in U(N/2)$.  In each case we find that the ansatz
(\ref{ansatz})--(\ref{eq:Uo}) yields a minimum that saturates the
lower bound (\ref{bound}).
Again, there is the possibility of accidental degeneracy.

\begin{table}[htb]
\caption{Breaking of $SU(N_f)_L\times SU(N_f)_R\times U(1)_A$ for
all baryon densities (per site) accessible for $N_f\le3$. Results
for $B=0$ are from \cite{paper1}. \label{table}}
\begin{ruledtabular}
\begin{tabular}{cccc}

$N_f$   &   $|B|$   &   Unbroken symmetry & Broken charges\\
\hline
    &   0   & $-$  & 1  \\
1   &   1   & $-$  & 1  \\
    &   2   & $U(1)_A$ & 0  \\  \hline
    &   0   & $SU(2)_V$ & 4  \\
    &   1   & $U(1)_{I_3}$ & 6 \\
2   &   2   & $SU(2)_V$ & 4  \\
    &   3   & $U(1)_{I_3}$ & 6  \\
    &   4   & $SU(2)_L\times SU(2)_R\times U(1)_A$ & 0  \\  \hline
    &   0   & $SU(3)_V$ & 9  \\
    &   1   & $U(1)_Y\times SU(2)_V$ & 13\\
    &   2   & $U(1)_Y$ & 16 \\
3   &   3   & $SU(3)_V$ & 9  \\
    &   4   & $U(1)_{I_3}\times U(1)_Y$ & 15 \\
    &   5   & $U(1)_{I_3}\times U(1)_Y\times U(1)_{A'}$ & 14  \\
    &   6   & $SU(3)_L\times SU(3)_R\times U(1)_A$ & 0 \\
\end{tabular}\end{ruledtabular}\end{table}

Upon calculating the $\sigma$ fields using
Eqs.~(\ref{ansatz})--(\ref{eq:Uo}), it is straightforward to
ascertain the symmetry of the vacuum. The $U(N_f)\times U(N_f)$
generators that commute with both $\sigma_e$ and~$\sigma_o$ form
the unbroken subgroup of the NNN theory. The rest are broken
generators that correspond to Goldstone bosons. We summarize our
results in Table \ref{table}, and we note the following:
\begin{itemize}
\item In 
cases of accidental degeneracy, we show the largest unbroken
symmetry attainable.  For $m\ge3N/4$ (i.e., $B\ge N_f$), this comes of the choice $u=\bm1_{N/2}$.
For $m<3N/4$ our numerical work cannot rule out vacua with yet larger symmetry.
\item Since the baryon background is fixed, we cannot tell whether
the $U(1)$ corresponding to baryon number is broken. [The $U(1)_B$
group acts trivially on $Q^\eta_\bn$ and on $\sigma_\bn$.] \item
For each value of $N_f$, the case $B=2N_f$ is a completely
saturated lattice.  Each site is in a singlet under the chiral
group, and there is no spontaneous symmetry breaking. \item The
axial $U(1)$ is not a symmetry of the continuum, and must be
broken by hand. Wherever it appears in Table \ref{table} it should
be neglected, whether as an unbroken symmetry or as a broken
charge. \item The $U(1)_{A'}$ appearing in the table for $N_f=3$
is not the original $U(1)_A$ group but rather is generated by
$\gamma_5 \otimes \lambda'$ with
\begin{equation}
\lambda'= \left( \begin{array}{ccc}
    1 & 0 & 0 \\
    0 & 0 & 0 \\
    0 & 0 & 0 \end{array} \right).
\end{equation}
This is the only case where an axial symmetry survives spontaneous
symmetry breaking. If $U(1)_A$ is broken by hand, so is
$U(1)_{A'}$.
\end{itemize}

For all nonzero densities short of saturation the vacuum breaks
rotational  invariance. This is easily checked for $m \ge 3N/4$ by
noting that the ansatz for $\sigma_{e,o}$ fails to commute with
the rotation operator~(\ref{R}). For the other cases this is
easily checked numerically. (In some cases a discrete symmetry
around the $z$ axis remains unbroken.) Since this is not a
continuous symmetry it will not give rise to additional Goldstone
bosons. The broken rotational invariance will of course affect the
excitation spectrum. In particular, whereas the NN theory
possesses excitations with linear and quadratic dispersion
relations \cite{odo}, the NNN theory will produce interesting
admixtures with anisotropic dispersion relations, like those seen
in \cite{Sannino}. We defer discussion of the excitations to a
future publication.

\section{Discussion}

In comparing our results to those of continuum CSC calculations,
one must keep in mind that we study systems with large, fixed, and
discrete values of $B$, rather than with large, continuous $\mu$.
Moreover, we use large-$N_c$ approximations which necessarily ignore the
discrete properties of the $SU(N_c)$ group that are essential to
baryons.  Quantum effects at finite $N_c$, treated correctly, should
yield effects that are not accessible through the $1/N_c$
expansion.

The values of $(N_c,N_f)$ that are of interest for CSC are $N_c=3$
and~$N_f=2$ or~3.  In the two-flavor case the favored $qq$
condensate is a flavor singlet and a color triplet, so that while color
is partially broken, chiral symmetry is unbroken.  We do not
see this for any density at $N_f=2$.  Plainly our results are due
to a $\bar qq$ condensate; whether there is a $qq$ condensate as
well cannot be ascertained.

For $N_f>2$ the situation is similar.  Sch\"afer \cite{Schafer}
has considered the color--flavor structure of the condensates that
arise for $N_c=3$ and $N_f\ge 3$, and he has found that both color
and flavor are partially broken, with a condensate that locks one
or more subgroups of the flavor group to the color group.  Since
we work at large $N_c$, we should stand the argument on its head.
A plausible $qq$ condensate would lock successive subgroups of the
{\em color\/} group to the flavor group and hence to each other,
leaving unbroken the diagonal $SU(N_f)_{L+R+C}$ and some leftover
color symmetry.  Judging by the global symmetry of the vacuum,
perhaps we see this for $(N_f=3,B=3)$.  The other cases could
conceivably arise from a combination of $\bar qq$ and~$qq$
condensates, but whether the latter actually occur is an open
question.

Finally we note that according to diagrammatic power-counting arguments
\cite{DGR,ShusterSon}, CSC should disappear in the 't~Hooft limit
($N_c\to\infty$, $g^2N_c$ fixed).  We do not strictly work in this
limit, since the large-$N_c$ approximation is applied only to the
effective strong-coupling Hamiltonian, after $g^2$ has disappeared
into setting the energy scale.

\begin{acknowledgments}
We thank Yigal Shamir for his assistance and Mark Alford for
valuable correspondence.  This work was supported by the Israel
Science Foundation under grant no.~222/02-1 and by the Tel Aviv
University Research Fund.
\end{acknowledgments}

\end{document}